\documentclass[structabstract]{aa}
\usepackage{natbib}                     % BibTeX style bibliography
\usepackage{amsmath}                    % Mathematic formulaes
\usepackage{graphicx}                   % Pictures and images
\usepackage[varg]{txfonts}              % A&A is printed using the Postscript TX Times-fonts.
\usepackage[english]{babel}             % Language is english
\usepackage{nameref}                    % References
\usepackage[normalem]{ulem}             % for striketrough type effects
\usepackage{paralist}                   % lists
\usepackage{multirow}
\def\kp{p}
\def\lp{k}

\usepackage{color}

\begin{document}

  \title{An extreme planetary system around HD\,219828\thanks{Based on observations collected with the HARPS spectrograph at the 3.6-m ESo telescope (La Silla-Paranal Observatory, Chile), runs ID 072.C-0488, 075.C-0332, 075.C-0332, 076.C-0155, 077.C-0101, 183.C-0972, 091.C-0936, and 192.C-0852, as well as with the ELODIE and SOPHIE spectrographs, at the OHP Observatory, France.}}
\subtitle{One long-period super Jupiter to a hot-neptune host star}

  \author{N. C. Santos\inst{1,2}
          \and A. Santerne\inst{1}
          \and J. P. Faria\inst{1,2}
          \and J. Rey\inst{3}
          \and A. C. M. Correia\inst{4,5}
          \and J. Laskar\inst{5}
          \and S. Udry\inst{3}
          \and V. Adibekyan\inst{1}
          \and F. Bouchy\inst{3,6}
          \and E. Delgado-Mena\inst{1}
          \and C. Melo\inst{7}
          \and X. Dumusque\inst{3}
          \and G. H\'ebrard\inst{8,9}
          \and C. Lovis\inst{3}
          \and M. Mayor\inst{3}
          \and M. Montalto\inst{1}
          \and A. Mortier\inst{10}
          \and F. Pepe\inst{3}
          \and P. Figueira\inst{1}
          \and J. Sahlmann\inst{11}
          \and D. S\'egransan\inst{3}
          \and S. G. Sousa\inst{1}
         }

  \institute{
          Instituto de Astrof\'isica e Ci\^encias do Espa\c{c}o, Universidade do Porto, CAUP, Rua das Estrelas, 4150-762 Porto, Portugal
          \and
          Departamento de F\'isica e Astronomia, Faculdade de Ci\^encias, Universidade do Porto, Rua do Campo Alegre, 4169-007 Porto, Portugal
          \and
          Observatoire de Gen\`eve, Universit\'e de Gen\`eve, 51 ch. des Maillettes, CH-1290 Sauverny, Switzerland
          \and
          CIDMA, Departamento de F\'isica, Universidade de Aveiro, Campus de Santiago, 3810-193, Aveiro, Portugal
          \and
          ASD, IMCCE-CNRS UMR8028, Observatoire de Paris, PSL Research University, 77 Av. Denfert-Rochereau, 75014 Paris, France
          \and
          Aix-Marseille Universit\'e, CNRS, Laboratoire d'Astrophysique de Marseille UMR 7326, 13388, Marseille Cedex 13, France
          \and
          European Southern Observatory (ESO), 19001 Casilla, Santiago, Chile
          \and
          Observatoire de Haute-Provence, CNRS/OAMP, 04870, Saint-Michel-l'Observatoire, France
          \and
          Institut d'Astrophysique de Paris, UMR7095 CNRS, Universit\'e Pierre \& Marie Curie, 98bis boulevard Arago, 75014 Paris, France 
          \and
          SUPA, School of Physics and Astronomy, University of St. Andrews, St. Andrews, KY16 9SS, UK 
          \and
          European Space Agency (ESA), Space Telescope Science Institute, 3700 San Martin Drive, Baltimore, MD 21218, USA
}

  \date{Received date / Accepted date }
%----------------------------------------------------------------------------------------
%       Abstract
%----------------------------------------------------------------------------------------
  \abstract
  {With about 2000 extrasolar planets confirmed, the results show that planetary systems have a whole range of unexpected properties. This wide 
  diversity provides fundamental clues to the processes of planet formation and evolution.}
  {We present a full investigation of the HD\,219828 system, a bright metal-rich star for which a hot neptune has previously been detected.}
  {We used a set of HARPS, SOPHIE, and ELODIE radial velocities to search for the existence of orbiting companions
  to HD\,219828. The spectra were used to characterise the star and its chemical abundances, as well as to check for spurious, activity induced signals.
 {A dynamical analysis is also performed to study the stability of the system and to constrain the orbital parameters and planet masses}. }
  {We announce the discovery of a long period (P=13.1\,years) massive ($m\,\sin{i}=15.1\,M_{Jup}$) companion (HD\,219828\,c) in a very eccentric ort 
  ($e=0.81$). The same data confirms the existence of a hot-neptune, HD\,219828\,b, with a minimum mass of 21\,M$_{\oplus}$ and a period of 3.83\,days. {The dynamical analysis shows that the system is stable, and that the equilibrium eccentricity of planet $b$ is close to zero.}}
  {The HD\,219828 system is extreme and unique in several aspects. First, ammong all known exoplanet systems it presents an unusually high mass ratio. 
  We also show that systems like HD\,219828, with a hot neptune and a long-period massive companion are more frequent than similar systems 
  with a hot jupiter instead. This suggests that the formation of hot neptunes follows a different path than the formation of their hot jovian counterparts.
  The high mass, long period, and eccentricity of HD\,219828\,c also make it a good target for Gaia astrometry as well as a potential target for atmospheric characterisation, using direct imaging or high-resolution spectroscopy. Astrometric observations will allow us to derive its real mass 
  and orbital configuration. If a transit of HD\,219828\,b is detected, we will be able to fully characterise the system, including the relative
  orbital inclinations. With a clearly known mass, HD\,219828\,c may become a benchmark object for the range in between giant planets and brown dwarfs. 
  }

%----------------------------------------------------------------------------------
%       Title
%----------------------------------------------------------------------------------

  \maketitle
%---------------------------------
   
%  -------------------------------------------------
%       Introduction
%----------------------------------------------------------------------------------
  \section{Introduction}                                        \label{sec:Introduction}

At a moment when the number of discovered extrasolar planets {is already above}
2000 \citep[see e.g.][http://www.exoplanet.eu]{Schneider-2011},
most of the attention in the field is given to detecting planets with increasingly lower mass
and to characterising their structure and atmospheres \citep[for some recent reviews
see e.g.][]{Mayor-2014,Burrows-2014,Lissauer-2014}.

While these efforts have mostly concentrated on short-period planets, which usually have a higher
detection probability (through transits and radial velocities) and can be more easily characterised, the detection of long-period giant planets still enjoys considerable
interest {\citep[see e.g.][]{Feng-2015,Santerne-2015,Moutou-2015}. } The frequency of systems with long-period giant planets \citep[e.g.][]{Rowan-2016}, which are more similar to
the solar system gas giants, may provide important clues to the formation of our
own system. Giant planets also play a crucial role in shaping the geometry of any
planetary system \citep[e.g.][]{Morbidelli-2007b}. Their detection thus provides important clues to the frequency and architecture
of different planetary systems, in particular concerning the existence of other Earth-like worlds, or concerning the formation process and properties of short-period planets \citep[e.g.][]{Nagasawa-2008,Izidoro-2015}.

In parallel, the formation of high-mass giant planets, in particular those on the borderline between giant planets and brown dwarfs, is still lively
debated. The distinction between the two types of objects has been argued to be
related to the deuterium-burning mass limit near 13\,M$_{Jup}$  \citep[][]{Burrows-2001}. The 
lack of a discontinuity in the mass distribution of companions to solar-type stars at the brown-dwarf
regime \citep[][]{Udry-2007,Sahlmann-2011} suggests, however, that a distinction based on the formation mechanisms may be more
representative \citep[][]{Chabrier-2014}. The detection and characterisation of high-mass planets in systems where lower
mass planets exist (e.g. the derivation of their masses, orbital properties, and relative inclinations) may in this sense provide 
relevant information.

\citet[][]{Melo-2007} presented the discovery of a short-period neptune-like planet 
orbiting the bright G0 sub-giant star \object{HD\,219828}. Evidence for a longer period signal in the data was
also presented, although no conclusive orbital solution was proposed. 
In this paper we present the results of the analysis of new radial-velocity measurements of this star, covering a time span of 16 years. 
The results allow us to fully confirm the detection of
the neptune-mass companion, but also to constrain the orbital solution and minimum mass
for the longer period counterpart.
In Sects.\,\ref{sec:observations} and \ref{sec:star} we present our new radial velocity dataset
and a comprehensive analysis of the stellar parameters and chemical abundances, respectively. 
The detected radial-velocity signals are discussed in Sect.\,\ref{sec:planet}, {and in Sect.\, \ref{bdsc} we present
a dynamical analysis of the system.} We then discuss
the relevance, uniqueness, and potential of this system in Sect.\,\ref{sec:conclusions}.

\section{Observations and data}                         
\label{sec:observations}

\citet[][]{Melo-2007} presented the discovery of a hot neptune orbiting HD\,219828
based on 22 high-precision radial velocities obtained with the HARPS
instrument at the 3.6 m ESO telescope \citep[La Silla-Paranal Observatory, Chile --][]{Mayor-2003b}. Since then,
69 new HARPS spectra were obtained with the goal of confirming the existence of
the short-period neptune but also to try to constrain the orbit of the longer period companion.
A total of 91 HARPS measurements over the same number of nights are now available,
spanning from 2005-05-19 to 2013-11-20. We note that two further measurements were
obtained in August 2015. However, a major instrument upgrade was made in June 2015 with the introduction of
octogonal fibers \citep[see][for the SOPHIE example and motivation]{Bouchy-2013}, and an offset in the radial velocities was introduced that is still not well characterised. As a result, these two 
new measurements do not allow us to add any relevant information. We thus decided to remove them 
from the current analysis.

The HARPS measurements were obtained with an exposure time of 900s and a simultaneous
calibration using the Fabry-P\'erot \'etalon mode whenever this was available. Otherwise, the traditional
ThAr simultaneous calibration mode was used for the older measurements that have been published in \citet[][]{Melo-2007}. This allowed
us to achieve an individual precision of better than 1\,m\,s$^{-1}$ for most of the measurements (the average
noise in our data is 80\,cm\,s$^{-1}$, including photon noise, calibration noise, and the uncertainty in the
measurement of the instrumental drift, with an rms of 20\,cm\,s$^{-1}$).
The exposure time used also allows averaging out the noise produced by stellar oscillation 
modes \citep[][]{Dumusque-2010}, with timescales of the order of a few minutes. This strategy does not allow us to completely overcome the granulation noise, however.

All HARPS spectra were reduced using the latest version of the HARPS pipeline. Precise
radial velocities were derived, together with measurements of the bisector inverse slope (BIS), full width at half maximum (FWHM),
and contrast of the HARPS cross-correlation function \citep[CCF --][]{Baranne-1996,Pepe-2002}. These values are
useful to diagnose radial-velocity variations that are intrinsic to the star, for instance caused by stellar
activity on its different timescales \citep[e.g.][]{Queloz-2000,Santos-2010a,Figueira-2010b,Dumusque-2012,Santerne-2015b}. Values for
the chromospheric activity index $\log{R'_{HK}}$ were also derived (see Sect.\,\ref{sec:star}) following a procedure
similar to the one adopted in \citet[][]{GomesdaSilva-2014} \citep[see also][]{Lovis-2011b}.

In addition to the HARPS data, radial-velocity measurements of HD\,219828 were also
obtained using the ELODIE \citep[][]{Baranne-1996} and SOPHIE \citep[][]{Perruchot-2008} cross-dispersor
fiber-fed echelle spectrographs, mounted on the 1.93 m telescope at the Observatoire de Haute Provence (OHP)\footnote{SOPHIE 
replaced ELODIE at the 193 cm telescope at the OHP observatory in France.}. Albeit with a lower precision when 
compared with HARPS, the radial velocities derived using these instruments are useful to constrain the period and eccentricity
of the long-period companion (see Sect.\,\ref{sec:planet}).

A total of four observations were made with ELODIE between 1999 and 2004 as part of the ELODIE planet search program \citep[][]{Perrier-2003}.
The first of the four observations from the ELODIE database was made without a simultaneous thorium-argon lamp but observed within the same mode immediately before the constant star HD\,220773. For this constant star, we determined the radial-velocity offset for observations obtained 
with and without a simultaneous thorium-argon lamp. We applied the same radial-velocity offset for HD\,219828. This allows us to be confident
of the measured radial velocity because the instrumental drift between the observations of the two stars is expected to be much lower than the
error bar of the measurement (of the order of 15-30 m\,s$^{-1}$).

A total of 19 measurements were also acquired with SOPHIE as part of the science verification at the end of 2007 and after the installation of octagonal fibers in 2011 \citep[][]{Bouchy-2013}. SOPHIE observations were secured in high-resolution mode and with thorium-argon simultaneous calibration. Exposure times were 900 or 1200 sec,
and the error bars of individual measurements have a value of 3.4\,m\,s$^{-1}$ on average.
Systematic effects were corrected in the SOPHIE measurements. All the spectra taken before June 2011 are corrected for the seeing effect \citep[][]{Boisse-2010,Boisse-2011b}. Spectra taken after June 2011 are corrected using a set of standard stars that are monitored every night. This correction accounts for any instrumental variations and is described in detail in \citet[][]{Courcol-2015}.

All the {used} radial velocities are made available in electronic form in CDS. A sample is provided in Table\,\ref{tab:rv}.

\begin{table}
\par
\caption{
\label{tab:rv}
Radial velocities (RV) for HD\,219828 together with the respective baricentric Julian dates, errors, and instrument used. }
\begin{tabular}{lccc}
\hline\hline
\noalign{\smallskip}
BJD  & RV & $\sigma(RV)$ & Instrument \\
$[-2400000\,days]$  & [$km\,s^{-1}$] & [$km\,s^{-1}$] & \\
\hline
51448.4592 &    $-$24.1328      &0.0300 &ELODIE\\
53281.4457&     $-$24.0628      &0.0150 &ELODIE\\
53332.3841&     $-$24.0828      &0.0150 &ELODIE\\
53334.3126&     $-$24.0628      &0.0150 &ELODIE\\
53509.927917&   $-$24.02444     &0.00090        &HARPS\\
53510.928367&   $-$24.01105     &0.00130 &HARPS\\
... & ... & ... & ...\\
\hline
\noalign{\smallskip}
\end{tabular}
\end{table}

\section{HD219828: the star}
\label{sec:star}

According to the revised version of the Hipparcos catalogue \citep[][]{vanLeeuwen-2007} and SIMBAD, \object{HD\,219828} is a G0IV star with a parallax $\pi$=12.83$\pm$0.74
mas, an apparent magnitude mv = 8.04, and a colour index $B-V$ = 0.654. These values imply,
adopting the bolometric correction from \citet[][]{Flower-1996}, an absolute magnitude Mv = 3.58 and a luminosity of
3.08$L_{\odot}$. Still according to Hipparcos, the star is stable up to 0.013\,mag (typical for a constant star of its
magnitude).

\begin{table}
\par
\caption{
\label{tab:star}
Stellar parameters for \object{HD\,219828}. }
\begin{tabular}{lcc}
\hline\hline
\noalign{\smallskip}
Parameter  & Value & Reference/Method \\
\hline
Spectral~type                   & G0IV                  & \citep[][]{vanLeeuwen-2007}  \\
Parallax~[mas]                  & 12.83 $~\pm~$1.01      & \citep[][]{vanLeeuwen-2007} \\
Distance~[pc]                   & 77.9                  & \citep[][]{vanLeeuwen-2007} \\
$m_v$                           & 8.04                  & \citep[][]{vanLeeuwen-2007} \\
$B-V$                           & 0.654                 & \citep[][]{vanLeeuwen-2007} \\
$M_{v}$                         & 3.58                 & -- \\
Bolometric correction & -0.061 & \citet[][]{Flower-1996}\\
Luminosity~$[L_{\odot}]$        & 3.08         & -- \\
Mass~$[M_{\odot}]$              & 1.23$\pm$0.10 & \citet[][]{Bressan-2012} \\
%Radius [R$_\odot$]     & 1.69 & -- \\
$\log{R'_{\rm HK}}$             & $-$5.12 & HARPS \\
$P_{rot}$ [days]             &      28.7/31.7          &  $(a)$\\
$v~\sin{i}$~[km~s$^{-1}$]       & 2.9   & HARPS\,$(b)$\\
$T_{\rm eff}$~[K]               & 5891$~\pm~$18         & \citet[][]{Melo-2007} \\
$\log{g}$                       & 4.08$~\pm~$0.10$^{(c)}$       & \citet[][]{Melo-2007} \\
$\xi_{\mathrm{t}}$              & 1.18$~\pm~$0.02       & \citet[][]{Melo-2007} \\
${\rm [Fe/H]}$                  & $+$0.19$~\pm~$0.03    & \citet[][]{Melo-2007} \\
\hline
\noalign{\smallskip}
\end{tabular}
\newline
$(a)$ Using the activity level and the calibrations of \citet[][]{Noyes-1984} and \citet[][]{Mamajek-2008}
\newline
$(b)$ Based on HARPS spectra using a calibration similar to the one presented by \citet{Santos-2002a}
\newline
$(c)$ Value after correction using the calibration of \citet[][]{Mortier-2014}
\end{table}

Using a high-resolution and high signal-to-noise ratio (S/N) HARPS spectrum, \citet[][]{Melo-2006} derived precise stellar atmospheric
parameters for this star \citep[see also SWEET-Cat --][]{Santos-2013}\footnote{http://www.astro.up.pt/resources/sweet-cat}. The values are presented in Table\,\ref{tab:star}. The derived effective temperature T$_{eff}$=5891$\pm$18\,K, $\log{g}$ = 4.19$\pm$0.05\,dex, and [Fe/H] = 0.19$\pm$0.02\,dex agree well with the
spectral type listed in Hipparcos and also with other values from the literature: \citet[][]{vanBelle-2009} derived a temperature 
of T$_{eff}$=5929$\pm$90 from SED fitting, and \citet[][]{Gonzalez-2010} derived T$_{eff}$=5861$\pm$38\,K and $\log{g}$=4.21\,dex. 

PARSEC evolutionary models \citep[][]{Bressan-2012}\footnote{See \citet[][]{LiciodaSilva-2006} -- http://stev.oapd.inaf.it/cgi-bin/param} indicate that HD\,219828 
has a mass of 1.23$\pm$0.06$M_{\odot}$ and an age of 4.6$\pm$0.7\,Gyr.
A similar age was found by \citet[][]{Casagrande-2011} (5$\pm$1.34\,Gyr), also compatible with its activity level \citep[][indicating an age above 2\,Gyr]{Pace-2013}.
A mass of 1.22$M_{\odot}$ is also found using the calibration of \citet[][]{Torres-2010} based on the spectroscopically
derived values of T$_{eff}$, $\log{g}$, and [Fe/H], after correcting for the systematics as explained in \citet[][]{Mortier-2014} and \citet[][]{Santos-2013}. 
We decided to adopt a value of 1.23$M_{\odot}$ to which we associated a conservative uncertainty of 0.10\,$M_{\odot}$.
A surface gravity value slightly lower than the one found through the spectroscopic analysis (4.09\,dex) was derived when we considered the above stellar mass, 
the bolometric correction, and the Hipparcos distance and magnitude following Eq. 1 of \citet[][]{Santos-2004b}. A similarly lower value (4.08\,dex) was also obtained when
we corrected the spectroscopic surface gravity using the calibration presented by \citet[][]{Mortier-2014}.
We therefore decided to adopt a lower surface gravity, with a conservative error bar of 0.10\,dex.

Our derived lithium (Li) abundance for HD\,219828 is $\log{\epsilon}(Li)$=2.33$\pm$0.04, similar to the value found by \citet[][]{Gonzalez-2010} ($\log{\epsilon}(Li)$=2.17$\pm$0.06), and typical of early-G dwarfs. Planet-host stars with a temperature close to solar have been suggested to be Li-poor when compared with single field dwarfs of the same age and metallicity \citep[for a recent paper and debate see][and references therein]{Figueira-2014}. However, the T$_{eff}$ of this star lies beyond the temperature range where this effect is observed (5700-5850K), and hotter stars show higher Li abundances regardless of the presence of planets \citep[e.g.][]{Delgado-Mena-2014}.

The derived radius of the star, based on the relation of luminosity,
temperature, and radius is 1.69\,R$_{\odot}$.
This value is slighly above the one derived using the \citet[][]{Torres-2010} calibration (1.47\,R$_{\odot}$), the spectral energy distribution
(SED) fitting procedure adopted in
\citet[][]{vanBelle-2009} (1.58$\pm$0.10\,R$_{\odot}$), or the one obtained from the PARSEC interface \citep[][1.61$\pm$0.12\,R$_{\odot}$]{LiciodaSilva-2006}.

In brief, HD\,219828 is a metal-rich early-G dwarf that is slightly evolved beyond the main sequence.
The adopted parameters for the star are listed in Table\,\ref{tab:star}.

HD\,219828 has also been found to be a low-activity star: from the HARPS spectra we derive a mean $\log{R'_{HK}}=-5.12$\,dex, with
a small dispersion of only 0.02\,dex. These values are even lower than those found in \citet[][]{Pace-2013} ($\log{R'_{HK}}$ between $-4.896$
and $-4.880$), attesting to the low activity level of this star. This low value for the activity level would correspond to a rotational period of
28.7$\pm$0.6\,days and 31.7$\pm$0.8\,days, according to the calibrations
of \citet[][]{Noyes-1984} and \citet[][]{Mamajek-2008}, respectively. 
A projected rotational velocity $v\,\sin{i}$=2.9\,km\,s$^{-1}$ is derived from the FWHM of the HARPS cross-correlation function \citep[see e.g.][]{Santos-2002a}\footnote{This calibration is made for dwarfs stars, however, and may not be fully valid for this slightly evolved star.}.
With a stellar radius of 1.69\,R$_\odot$ and assuming a rotational period of 30\,days, HD\,219828 should be rotating with V$_{rot}$=2.8\,\,km\,s$^{-1}$.
These values suggests that the star may be seen almost equator-on.
                        
The abundances of planet-host stars as well as their origin in the Galaxy have been suggested to have an important
influence on the frequency, composition, architecture, and formation history \citep[][]{Adibekyan-2012,Dawson-2013, Adibekyan-2013,Santos-2015a}.
We therefore decided to derive detailed abundances for several $\alpha$- and iron-peak elements in HD\,219828 
using a combined HARPS spectrum, built from all the data available at the time of this publication\footnote{The combined spectrum has a S/N above 1000 per pixel.}.
The method we adopted is fully explained in \citet[][]{Adibekyan-2012b}, and we refer to that paper for more details. The stellar atmospheric
parameters adopted for the analysis are those listed in Table\,\ref{tab:star}. The final set of
abundances, together with the number of element lines used to derive the abundances and respective errors (denoting the
error on the mean) are listed in Table\,\ref{tab:abundances}. Moreover, we also derived abundances of several volatile elements (C, O, S, and Zn) and heavier elements, 
following the procedures of \citet[][]{Ecuvillon-2004b} and \citet[][]{BertrandeLis-2015}. The values allow us to conclude that HD\,219828 is
a typical thin-disk star from the solar neighbourhood; its [$\alpha$/Fe]\footnote{Where $\alpha$-elements are e.g. Ti, Si, and Mg, and the notation reads as 
[X/Y]=[X/H]-[Y/H]} abundance ratios are nearly solar (i.e. $\sim$0.0): no $\alpha$-element enhancement, typical of thick-disk or halo stars, 
is observed \citep[see][and references therein]{Adibekyan-2012b}.

\begin{table}
\par
\caption{
\label{tab:abundances}
Chemical abundances for \object{HD\,219828}. }
\begin{tabular}{lcl}
\hline\hline
\noalign{\smallskip}
Element  (X) & [X/H] & N$_{lines}$ \\
\hline
CI & 0.093$\pm$0.002 & 2 \\
OI & 0.188$\pm$0.047 & 2 \\
NaI & 0.191$\pm$0.015 & 2 \\
MgI & 0.180$\pm$0.030 & 3 \\
AlI & 0.204$\pm$0.009 & 2 \\
SiI & 0.177$\pm$0.018 & 14 \\
SI & 0.085$\pm$0.017 & 4 \\
CaI & 0.165$\pm$0.027 & 12 \\
ScI & 0.252$\pm$0.028 & 3 \\
ScII & 0.254$\pm$0.015 & 6 \\
TiI & 0.191$\pm$0.025 & 21 \\
TiII & 0.173$\pm$0.040 & 5 \\
VI & 0.196$\pm$0.010 & 8 \\
CrI & 0.153$\pm$0.014 & 18 \\
CrII & 0.116$\pm$0.034 & 3 \\
MnI & 0.181$\pm$0.035 & 5 \\
CoI & 0.216$\pm$0.086 & 8 \\
NiI & 0.191$\pm$0.017 & 40 \\
CuI & 0.238$\pm$0.019 & 4 \\
ZnI & 0.158$\pm$0.005 & 3 \\
SrI & 0.103$\pm$0.050 & 1 \\
YI & 0.185$\pm$0.070 & 7 \\
ZrI & 0.106$\pm$0.040 & 4 \\
BaI & 0.138$\pm$0.031 & 3 \\
CeI & 0.122$\pm$0.052 & 4 \\
NdI & 0.095$\pm$0.057 & 2 \\
\hline
A(Li) & 2.33 $\pm$0.03 & 1 \\
\hline
\noalign{\smallskip}
\end{tabular}
\end{table}

\section{Radial-velocity fitting}
\label{sec:planet}

\begin{figure}
\begin{center}
\begin{tabular}{c}
\includegraphics[angle=0,width=1\linewidth]{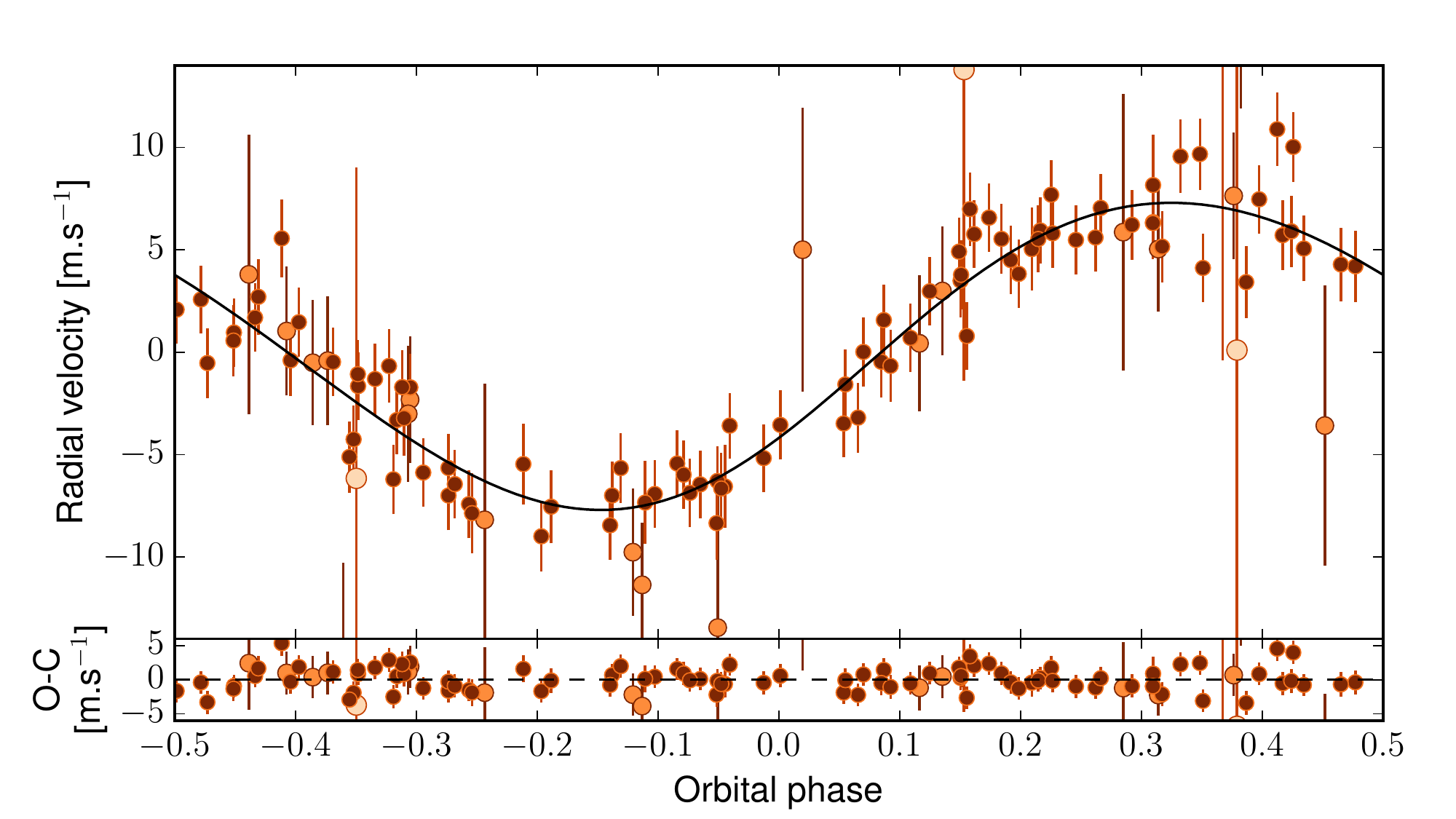}\\
\includegraphics[angle=0,width=1\linewidth]{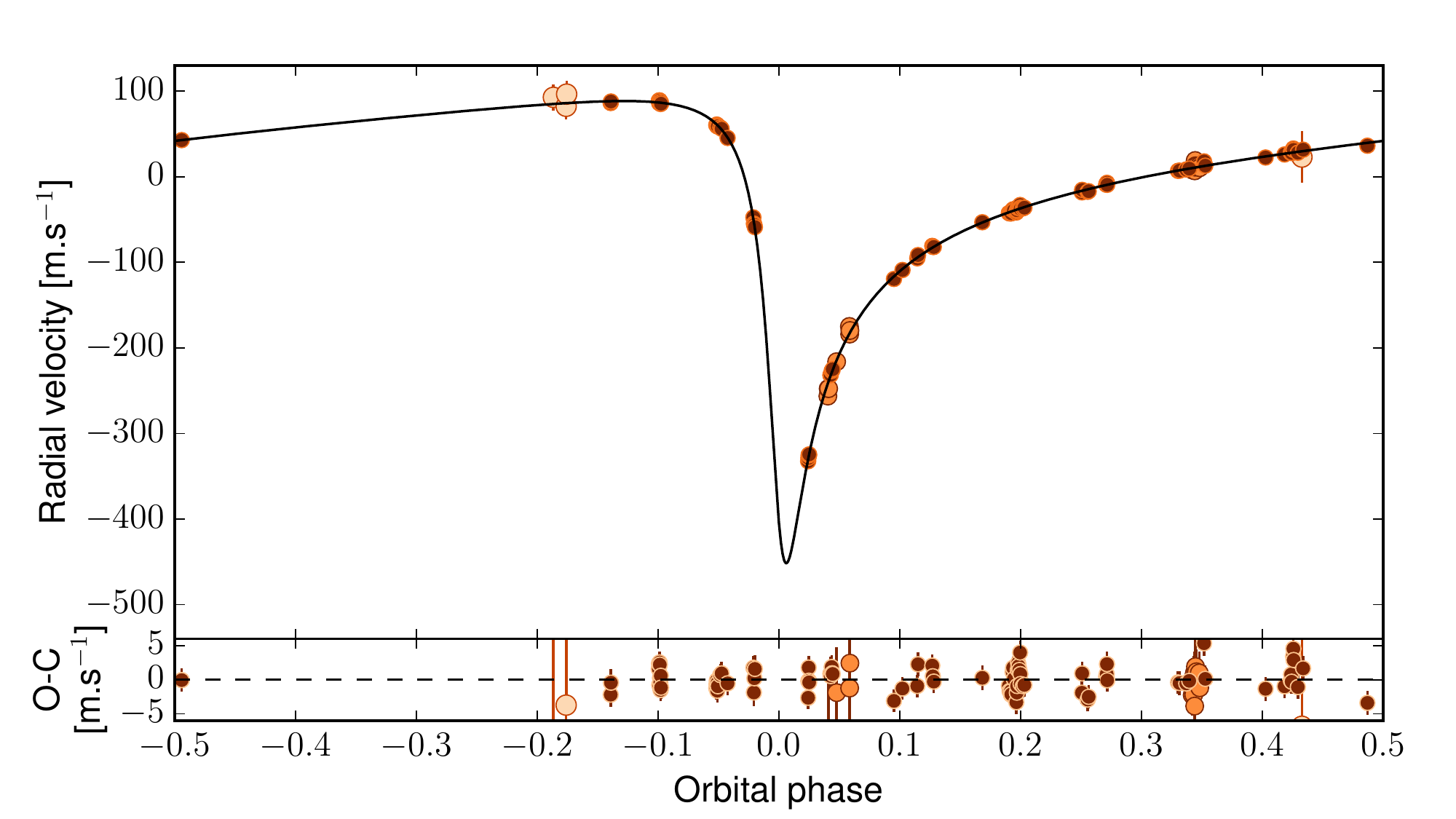}\\
\includegraphics[angle=0,width=1\linewidth]{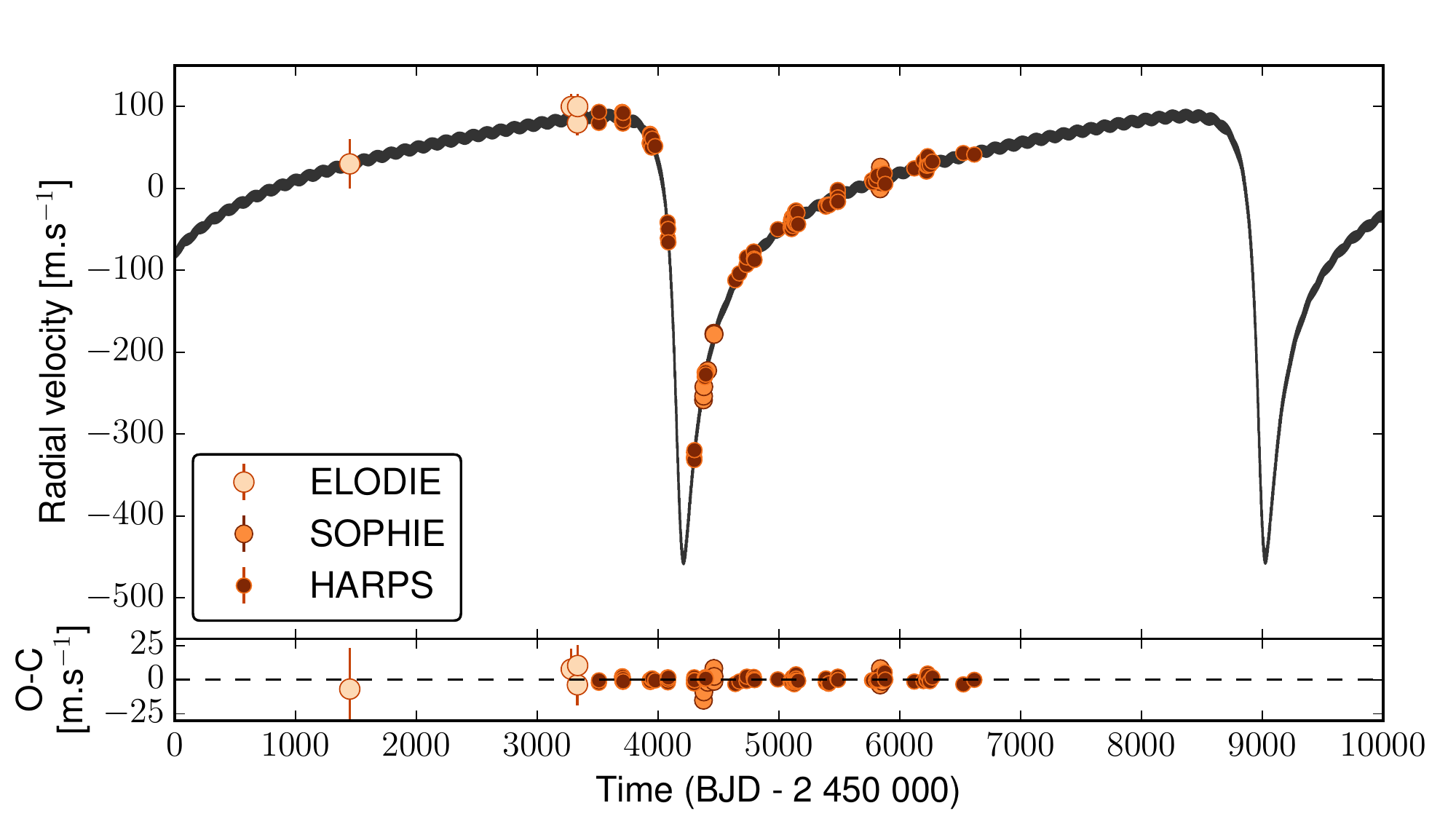}
\end{tabular}
\end{center}
\vspace{-0.5cm}
\caption{Top and middle plots: phase-folded radial-velocity curves and data for planets b and c, respectively. 
Bottom: combined orbital solution of planets b and c as a function of time. Residuals of the complete fit
are presented at the bottom of each diagram.}
\label{fig:orbits}
\end{figure}

The HARPS, SOPHIE, and ELODIE data were fit simultaneously using the Markov chain Monte Carlo (MCMC) algorithm implemented into the PASTIS software and fully described in \citet[][]{Diaz-2014}.
The priors used are listed in Table\,\ref{tab:priors}. The number of Keplerian functions was set to two: one to take the short-period signal into account that has been announced in \citet[][]{Melo-2007}, and the other to fit the long-term signal that is clear by visual inspection of
the data (see also Fig.\,\ref{fig:orbits}). The offsets between the HARPS, SOPHIE, and ELODIE radial velocities were also
fit during the process.

\def\ms{\,m\,s$^{-1}$}         %m.s -1
\def\kms{\,km\,s$^{-1}$}         %m.s -1
\begin{table*}[]
\caption{List of free parameters and respective priors used in the MCMC analysis of the radial velocities. }
\begin{center}
\begin{tabular}{lcc}
\hline
\hline
Parameter & \multicolumn{2}{c}{Prior}\\
\\
\hline
& &\\
& Planet b & Planet c\\
& &\\
Orbital period $P$ [d] & $\mathcal{J}(0.3;100)$ & $\mathcal{J}(100;1\times10^{5})$\\
Epoch of periastron T$_{p}$ [BJD] & $\mathcal{U}(2453000;2453200)$ & $\mathcal{U}(2450000;2650000)$\\
Orbital eccentricity $e$ & $\beta(0.867;3.03)$ & $\beta(0.867;3.03)$ \\
Argument of periastron $\omega$ [\degr] & $\mathcal{U}(0;360)$ & $\mathcal{U}(0;360)$ \\
Radial-velocity amplitude $K$ [\ms] & $\mathcal{U}(0;1000)$ & $\mathcal{U}(0;1000)$\\
\hline
& &\\
Systemic radial velocity $\gamma$ [\kms] & \multicolumn{2}{c}{$\mathcal{U}(-100, 100)$}\\
ELODIE radial-velocity jitter [\ms] & \multicolumn{2}{c}{$\mathcal{U}(0, 200)$}\\
SOPHIE radial-velocity jitter [\ms] & \multicolumn{2}{c}{$\mathcal{U}(0, 200)$}\\
HARPS radial-velocity jitter [\ms] & \multicolumn{2}{c}{$\mathcal{U}(0, 100)$}\\
ELODIE -- HARPS radial-velocity offset [\kms] & \multicolumn{2}{c}{$\mathcal{U}(-2, 2)$}\\
SOPHIE -- HARPS radial-velocity offset [\ms] & \multicolumn{2}{c}{$\mathcal{U}(-200, 200)$}\\
\hline
\hline
\end{tabular}
\tablefoot{$\mathcal{J}(a;b)$ is a Jeffreys distribution between $a$ and $b$, $\mathcal{U}(a;b)$ is a uniform distribution between $a$ and $b$, and $\beta(a;b)$ is a beta distribution with parameters $a$ and $b$.}
\tablebib{The choice of prior for the orbital eccentricity is described in \citet{Kipping-2013}.}
\end{center}
\label{tab:priors}
\end{table*}%

We ran 100 chains of $3\,10^5$ iterations starting from an initial value randomly drawn from the joint prior. We then removed chains that did not converge to the 
maximum likelihood. We also removed the burn-in before thinning and merging the remaining chains. This provided 
10\,000 independent samples of the posterior distribution.

{The best-fit parameters of the orbits and their 68.3\% confidence interval are listed in Table\,\ref{tab:orbit}. The best
solutions hold for planets $b$ and $c$, periods of 3.83 and 4791\,days, eccentricities of
0.059$\pm$0.036 and 0.8115$\pm$0.0032, and planet masses of 21$\pm$1.4\,M$_\oplus$ and 15.1$\pm$0.85\,M$_{Jup}$. The eccentricity value found
for HD\,219828\,b is not significant at a 1.6 sigma level, and we therefore consider that we have no solid ground
to assume that it is different from zero. On the other hand, the high value for the eccentricity for HD\,219828\,c
brings this companion to as close as 1.13\,AU from its host star at periastron, while at apastron the
distance is around 10.8\,AU}.

We also tested the possibility that an additional long-term drift is present in the data. Adding this additional term to 
the fitting procedure did not return any significant result. We exclude any linear drift with an amplitude higher 
than 1\,m\,s$^{-1}$\,yr$^{-1}$ with 95\% confidence.

\begin{table*}
\par
\begin{center}
\caption{
\label{tab:orbit}
Orbital parameters for the two Keplerian solutions and the derived planet minimum mass.}
\begin{tabular}{lcc}
\hline\hline
\noalign{\smallskip}
$\gamma$ [km\,s$^{-1}$]         &       \multicolumn{2}{c}{$-$24.1047$\pm$0.0013}\\
\hline
& & \\
& \multicolumn{1}{c}{HD219828\,b} & \multicolumn{1}{c}{HD219828\,c}\\
& & \\
K [m\,s$^{-1}$]                         & 7.53$\pm$0.28         & 269.4$\pm$4.7 \\
P [days]                                & 3.834887$\pm$0.000096 & 4791$\pm$75 \\ 
Tp [BJD-2400000] & 55998.78$\pm$0.35            & 54180.7$\pm$1.2\\
e                                       & 0.059$\pm$0.036       & 0.8115$\pm$0.0032\\
$\omega$        [$\deg$]                & 225$\pm$38                    & 145.77$\pm$0.28  \\
m$_2$\,$\sin{i}$ &      $21.0\pm\,1.4\,M_{\oplus}$                      & $15.1\pm0.85\,M_{Jup}$  \\
a [AU] & 0.045 & 5.96 \\
\hline
& & \\
HARPS jitter  [m\,s$^{-1}$]                             & \multicolumn{2}{c}{1.64$\pm$0.17} \\
SOPHIE jitter  [m\,s$^{-1}$]                            & \multicolumn{2}{c}{2.75$\pm$1.10} \\
ELODIE jitter  [m\,s$^{-1}$]                            & \multicolumn{2}{c}{11.4$^{+20}_{-8.3}$} \\
\hline
& & \\
SOPHIE offset (relative to HARPS)  [m\,s$^{-1}$]  & \multicolumn{2}{c}{3.06$\pm$0.96} \\
ELODIE offset (relative to HARPS)  [m\,s$^{-1}$]  & \multicolumn{2}{c}{54$\pm$12}\\
\hline
\noalign{\smallskip}
\end{tabular}
\end{center}
\end{table*}

\begin{figure*}[t!]
%\begin{center}
\begin{tabular}{lr}\\
\includegraphics[width=9.1cm]{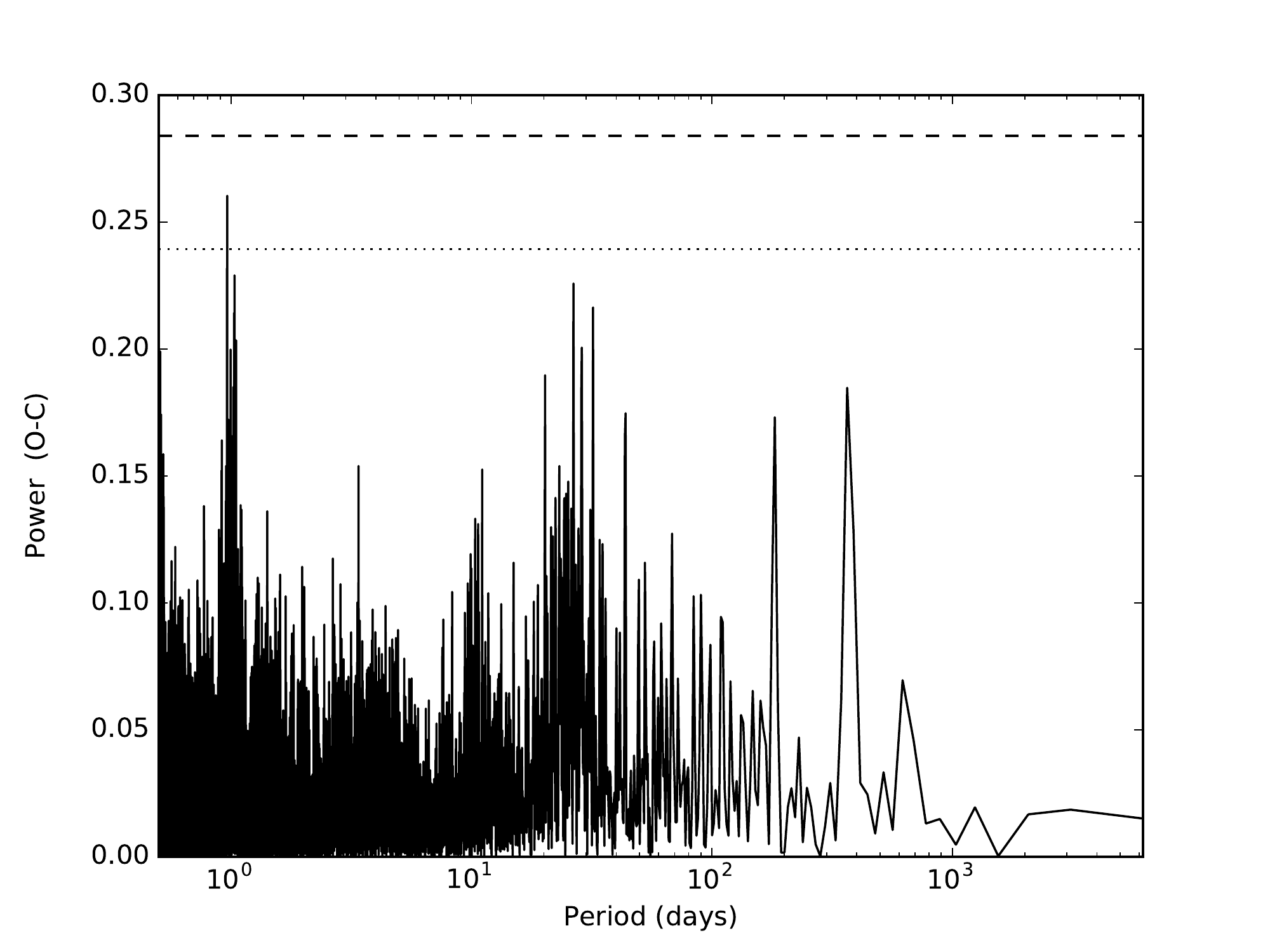} &
\includegraphics[width=9.1cm]{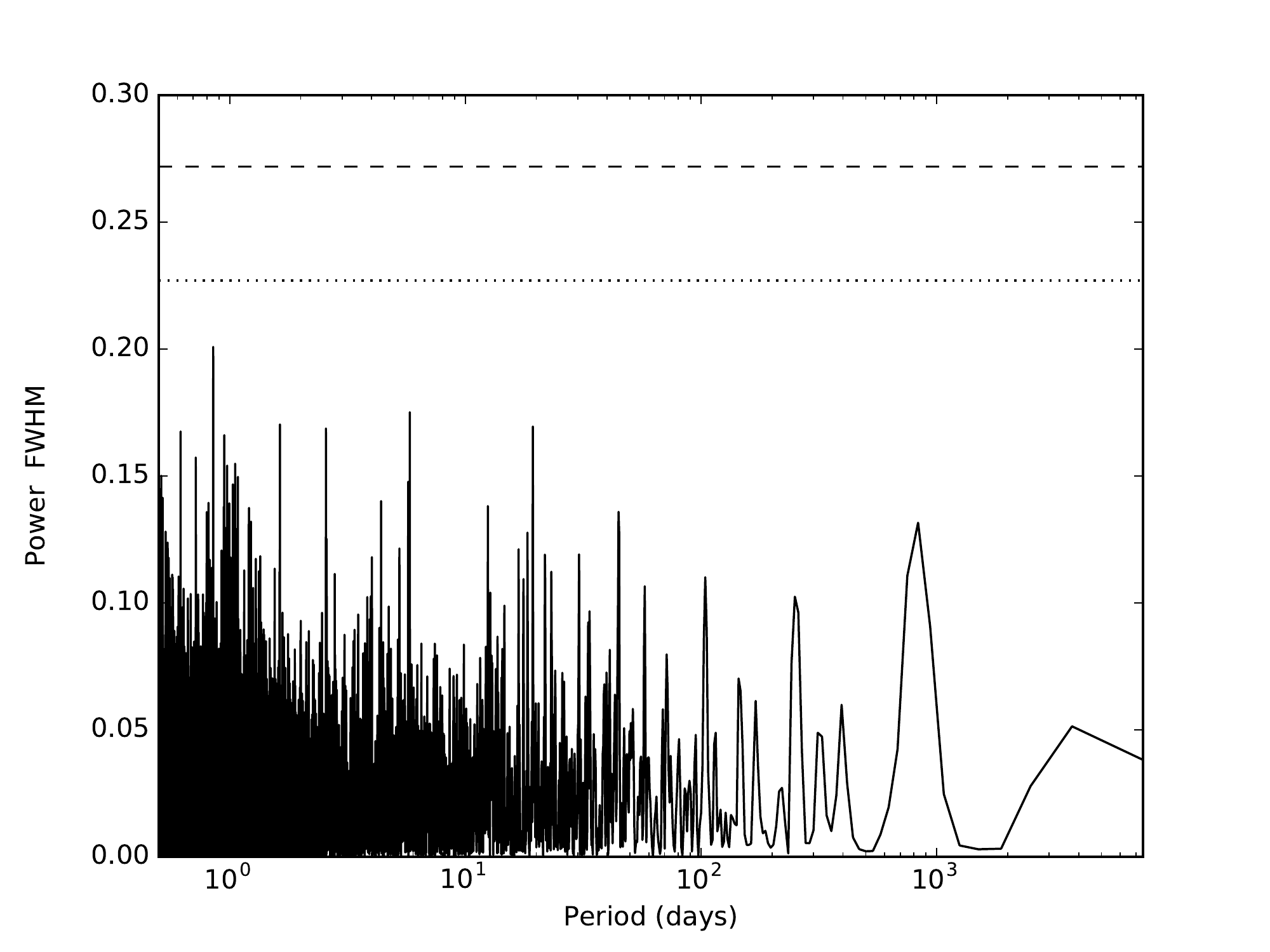}\\
\includegraphics[width=9.1cm]{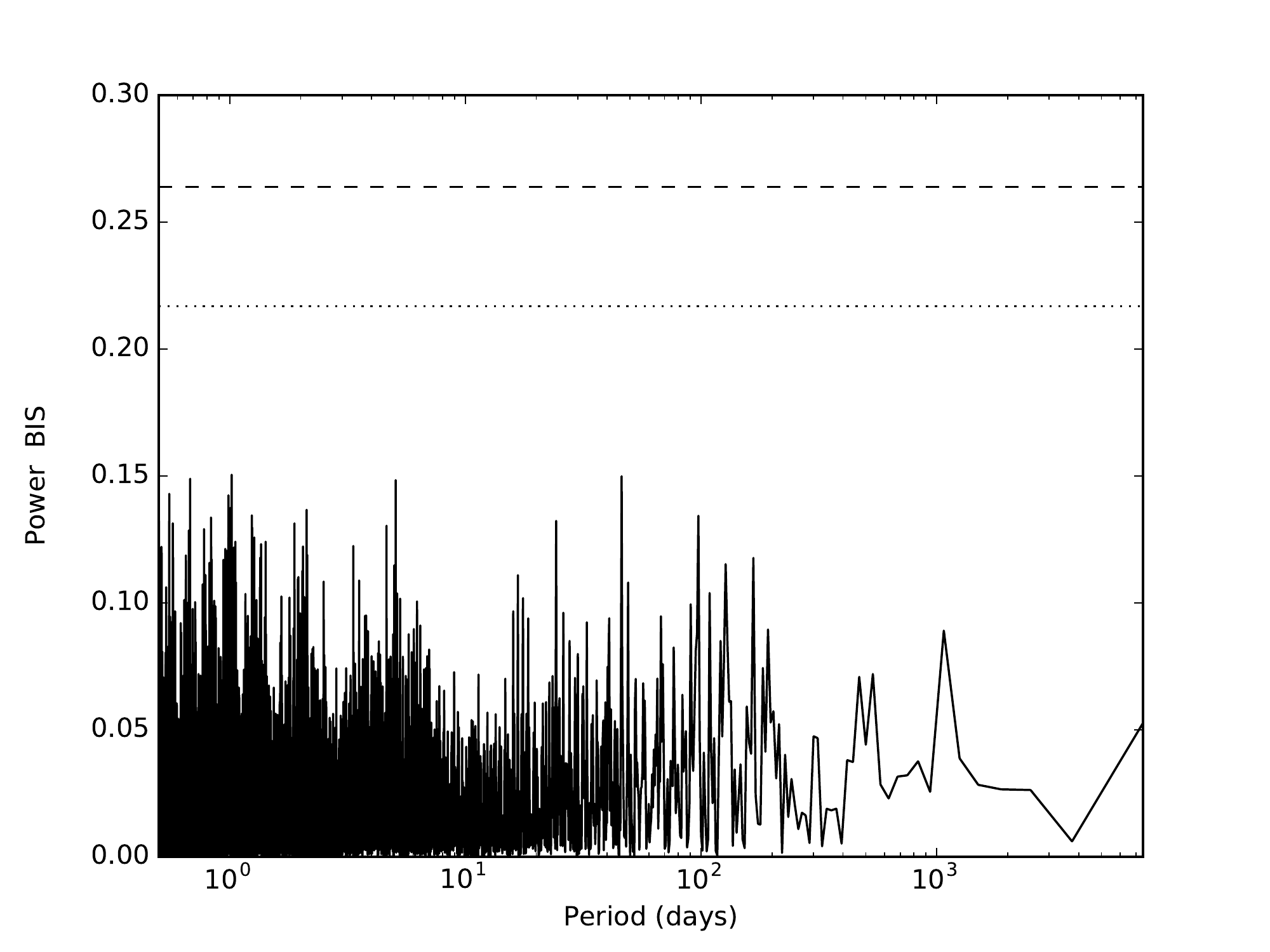}&
\includegraphics[width=9.1cm]{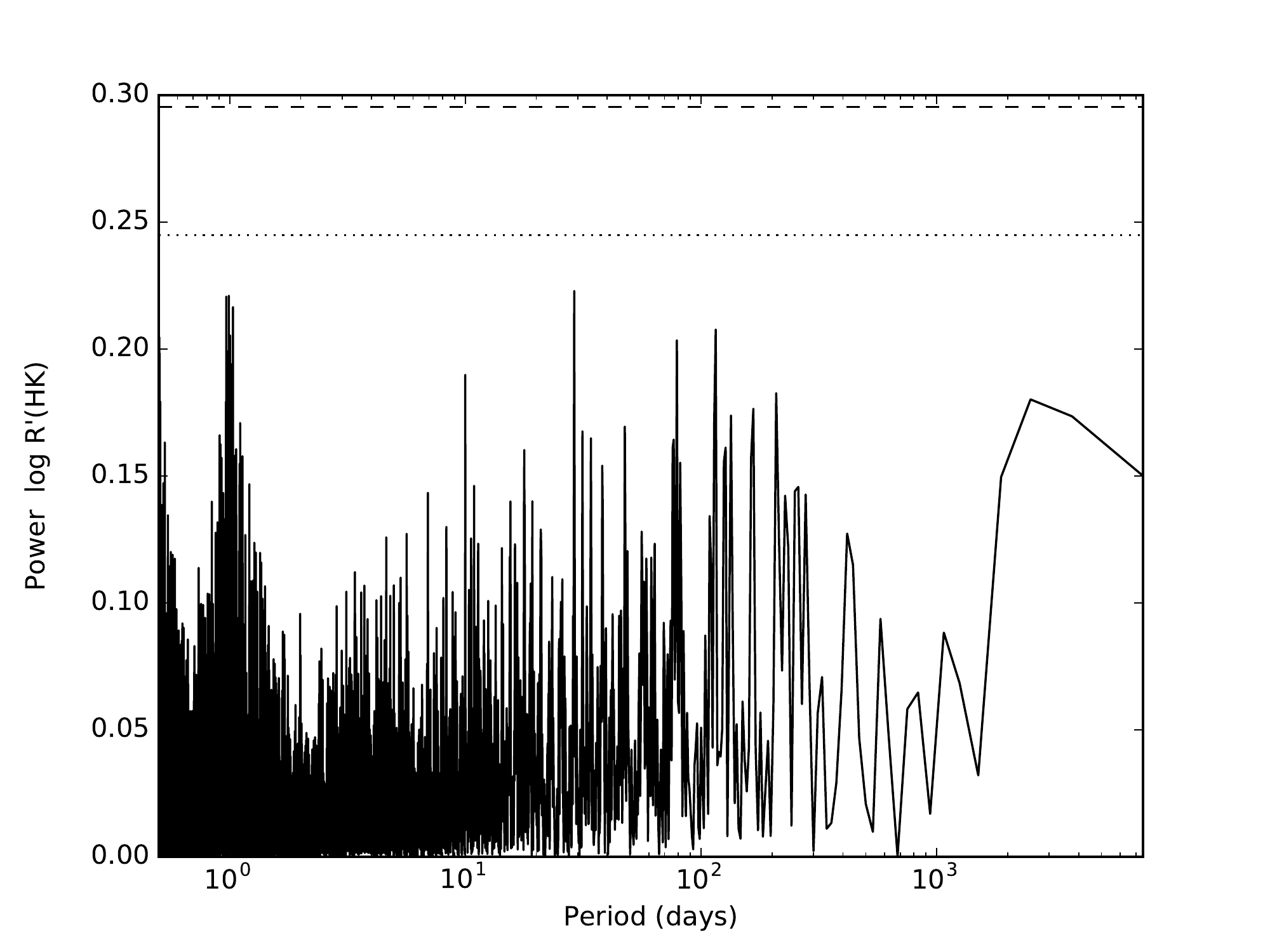}\\
\end{tabular}
%\end{center}
%\vspace{-0.5cm}
\caption{Periodograms of the radial-velocity residuals to the two-Keplerian fit
and of the raw values of FWHM, BIS, and $\log{R'_{HK}}$. The dotted and dashed horizontal lines denote
the 10\% and 1\% false-alarm probability levels, respectively.}
\label{fig:periodograms}
\end{figure*}

Table\,\ref{tab:orbit} shows that the estimated jitter of the HARPS radial velocities is of the order of 1.64\,m\,s$^{-1}$,
significantly above the average error bar of the individual measurements of 0.8\,m\,s$^{-1}$. This difference may be caused by
the fact that HD\,219828 is an early-G star, slightly evolved beyond the main sequence. It is indeed well known
that stars of earlier type and more evolved stars present higher granulation noise in radial velocities \citep[see][and references therein]{Dumusque-2010}.
It is also known that stellar activity can produce signals that can mimic planetary companions at
different timescales {\citep[e.g.][]{Saar-1997,Queloz-2000,Santos-2010a,Dumusque-2012,Kane-2016}}, even for stars with a low activity level \citep[e.g.][]{Santos-2014}. 
{The analysis of the generalized Lomb Scargle (GLS) periodograms of the residuals (O$-$C) of radial velocities after removing the two Keplerians reveals some 
power near 30 days (Fig.\,\ref{fig:periodograms}, top left panel). This signal, although not statistically significant, is most
likely related to 
the rotational period of the star, which is estimated to be close to that value (Table\,\ref{tab:star}).}
To diagnose any activity-induced signal in Fig.\,\ref{fig:periodograms}, we also present the GLS periodogram of the FWHM, BIS, and $\log{R'_{HK}}$ time series. 
A peak at a period of 30\,days is also found in the periodogram of the $\log{R'_{HK}}$ (lower right), again not
significant. In all the plots, the dotted and dashed horizontal lines denote
the 10\% and 1\% false-alarm probability levels, respectively. These were derived using a permutation
test as done in \citet[][]{Mortier-2012}. No significant signal appeared in any of the variables.
We therefore have no evidence that stellar activity may be contributing to produce
the observed signals or any other additional signal in the data, at any timescale.
The lack of any long-term significant variability on the BIS and FWHM also excludes the possibility that HD\,219828\,c is a
solar-like stellar companion in a highly inclined orbit \citep[][]{Santerne-2015b}, since such a companion is expected to 
leave a trace on these parameters.

We also tested whether this extra noise might indicate another low-amplitude companion to the
system. No satisfactory Keplerian fit was found. We therefore find no compelling evidence for additional companions.

In brief, we find clear evidence for the presence of two companions orbiting HD\,219828: a neptune-mass
planet in a short-period circular orbit, and a high-mass companion on the borderline between a giant planet and brown dwarf
orbiting in a long-period eccentric trajectory. This result fully confirms the detection of \citet[][]{Melo-2007}, and the fitted parameters are compatible with those listed in the announcement paper within the error bars. We note that using the new dataset
alone also leads to an orbital solution for
the short-period planet that is compatible with the one presented in Melo et al. We did not detect significant eccentricity for the inner planet either,
with an upper limit of 0.15 within a 99\% confidence interval.

\begin{figure}[t]
\begin{center}
\begin{tabular}{c}
\includegraphics[angle=0,width=1\linewidth]{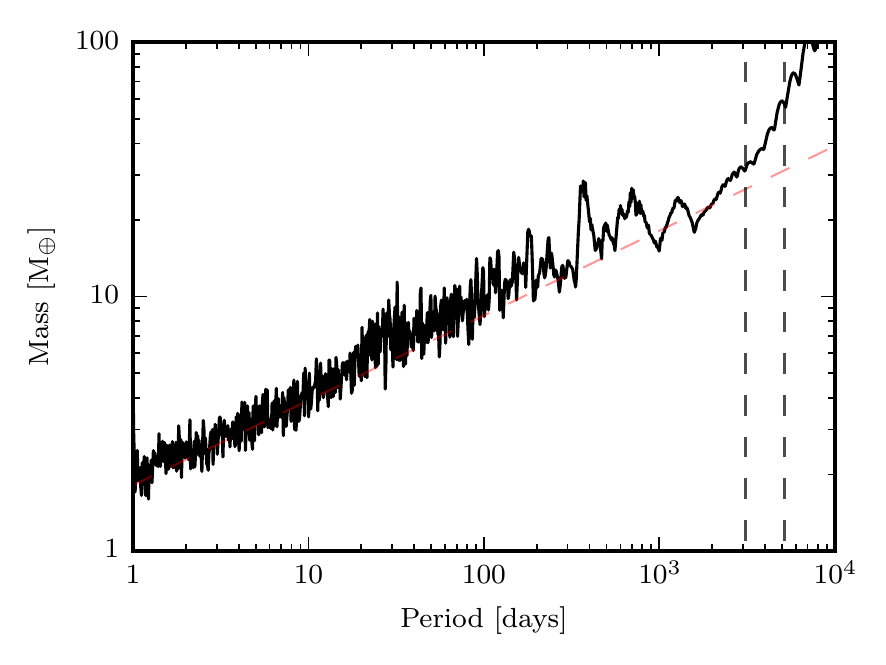}\\
\end{tabular}
\end{center}
\vspace{-0.5cm}
\caption{Detection limits for additional planets in the system.}
\label{fig:limits}
\end{figure}

We also computed the detection limits for additional planets in the system. For this
we analysed the residuals of the fit presented above, using a Monte Carlo approach similar to the one used in \citet[][]{Mortier-2012}. In brief, for each period we injected signals in the best-fit residuals, assuming circular orbits, with varying amplitudes
until the false-alarm probability of the detection was higher than 1\%. The amplitude of
this signal (for each period) sets the mass of the planet that can be detected
within the existing data.  The results of this analysis are presented in Fig.\,\ref{fig:limits}.
In the figure, the two vertical lines denote the time span of the HARPS observations and the total time span 
including the older ELODIE and SOPHIE data. The red line corresponds to the detection limit when we were able to detect a signal corresponding
to a circular orbit with a semi-amplitude of 1\,m\,s$^{-1}$. The results of this analysis suggest 
that we can exclude additional planets with masses above 10\,M$_\oplus$ in the
period range up to $\sim$100 days. This value decreases to 4\,M$_\oplus$ when
we restrict the period range to shorter than 10 days, and increases to $\sim$20\,M$_\oplus$
at periods of 1000\,days.
                                        
%______________________________________________________________
%
\section{Dynamical analysis}
%______________________________________________________________
\label{bdsc}

{ 

The orbital solution given in Table~5 shows a planetary system composed of two planets in a very uncommon configuration: a neptune-mass planet in a compact nearly circular orbit ($a_b = 0.05$~AU, $e_b=0.06$), together with a brown-dwarf-mass planet in a wide very eccentric orbit ($a_c = 6.0$~AU, $e_c=0.81$).
The stability of the system is not straightforward to gauge since the pericenter of the massive outer planet is nearly at 1~AU.
Gravitational perturbations on the inner planet cannot be neglected and may give rise to some instability.

\subsection{Stability analysis}
\label{saBD}

\begin{figure}
  \centering
    \includegraphics*[width=\columnwidth]{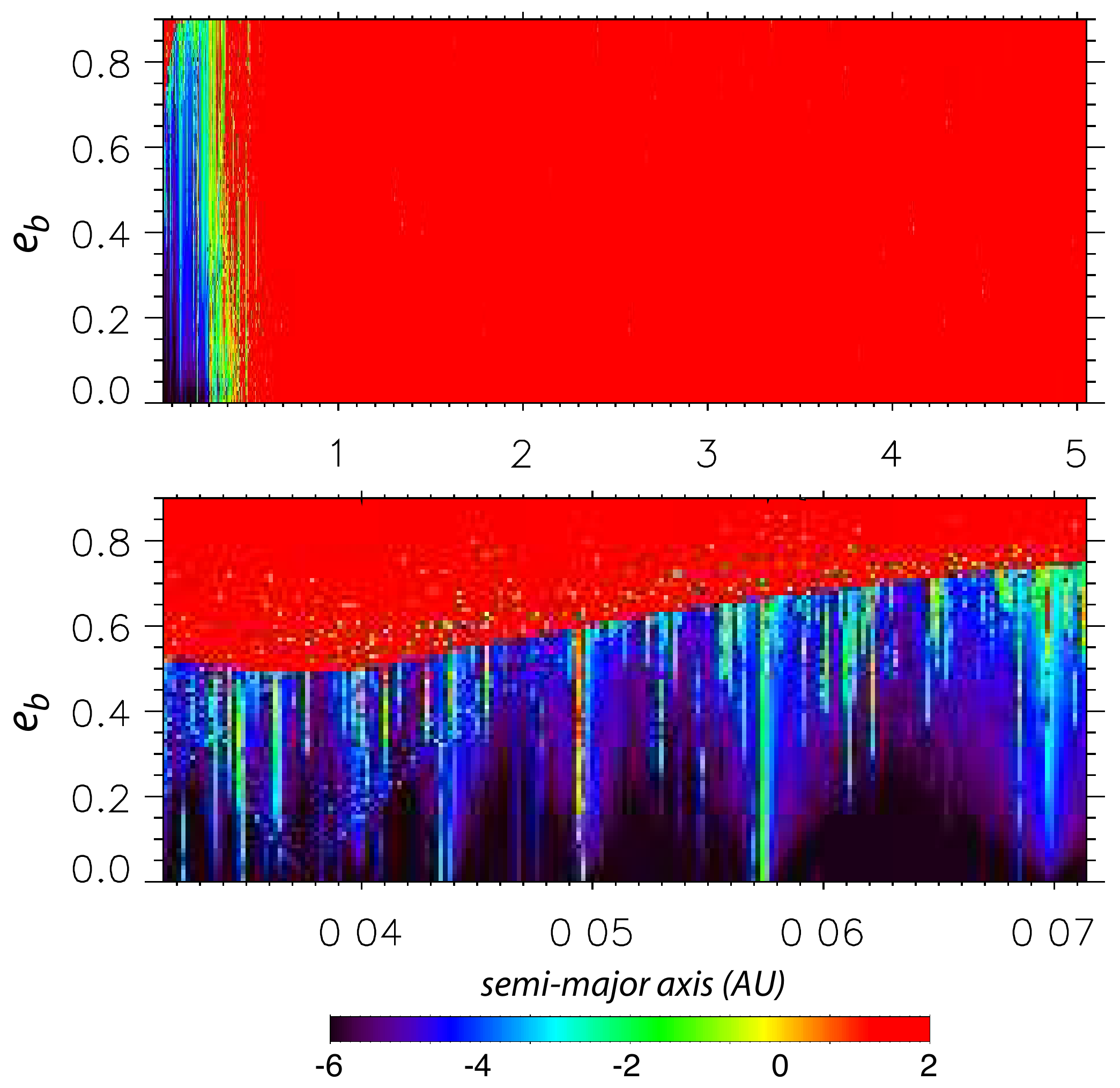} 
  \caption{
  Stability analysis of the nominal fit of the HD\,219828 planetary system (Table~5). 
  For fixed initial conditions, the phase space of the system is explored by varying the semi-major axis $a_b$ and eccentricity $e_b$ of the inner planet.
  The step size is $10^{-2}$ in eccentricity, and $5\times 10^{-3}$ (top) or $2\times 10^{-4}$~AU  (bottom) in semi-major axis. 
  For each initial condition, the system is integrated over $10^3$\,yr and a stability criterion is derived  with the frequency analysis of the mean longitude \citep{Laskar_1990,Laskar_1993PD}.
  As in \citet{Correia_etal_2005, Correia_etal_2009, Correia_etal_2010}, the chaotic diffusion 
  is measured by the variation in the frequencies. 
  The red zone corresponds to highly unstable orbits, while  the dark blue region can be assumed to be stable on a billion-year timescale.
  \label{Fdyn3}}   
\end{figure}

To analyse the stability of the nominal solution (Table~5), we performed a global frequency analysis \citep{Laskar_1993PD} in the vicinity of this solution, in the same way as achieved for other planetary systems \citep[e.g.][]{Correia_etal_2005,Correia_etal_2009,Correia_etal_2010}.
For each planet, the system is integrated on a regular 2D mesh of initial conditions, with varying semi-major axis and eccentricity, while the other parameters are retained at their nominal values (Table~5). 
The solution is integrated over 200~yr for each initial condition and a stability indicator is computed to be the variation in the measured mean motion over the two consecutive 100~yr intervals of time \citep[for more details see][]{Correia_etal_2005}.
For regular motion, there is no significant variation in the mean motion along the trajectory, while it can vary significantly for chaotic trajectories. 
The result is reported using a colour index in Fig.~\ref{Fdyn3}, where red represents the strongly chaotic trajectories and dark
blue  the extremely stable ones. 

In Fig.\,\ref{Fdyn3} we show the wide vicinity of the best-fit solution of the inner orbit.
We observe that from $0.2$ to 5~AU the system is totally unstable as a result of the presence of the outer planet (top).
However, when we zoom into the region $a<0.1$~AU (bottom), we verify that planet $b$ is stable even for eccentricities {up to 0.5}.
This figure thus show us that the HD\,219828 planetary system listed in Table~5 is stable over a timescale of several Gyr.

We can also try to constrain the highest possible masses of the planets if we assume co-planarity of the orbits.
By decreasing the inclination of the orbital plane of the system, we increase the mass values of {both} planets. We repeated a stability analysis of the orbits,
like in Fig.~\ref{Fdyn3}.
With decreasing inclination, the stable dark- blue areas become narrower,
to the point that the best-fit solution lies outside the stable zones.
At this point, we conclude that the system cannot be stable anymore.
It is not straightforward to find a transition inclination between the two regimes, but our analysis suggests that stability of the whole system is still possible for an inclination of $5^\circ$, but becomes impossible for lower values. 
Therefore, we conclude that the highest masses of the planets correspond to a scaling factor of about $10$, which would transform the inner planet into a hot Jupiter and the outer one into an M-dwarf star.

\subsection{Secular coupling}

\begin{table}
 \caption{Fundamental frequencies for the nominal orbital solution in
 Table~5. $n_b$ and $n_c$ are the mean motions, and $g_1$ and $g_2$ are the
 secular frequencies of the pericenters.
 \label{Tdyn1}} 
 \begin{center}
 \begin{tabular}{ccccr}
 \hline\hline
 \multicolumn{2}{c}{Frequency ($^\circ$/yr)}   & Period (yr) & \multicolumn{2}{c}{Angle (deg)} \\
 \hline
 $n_b$ & $3.428779 \times 10^{+4}$ & $1.049936 \times 10^{-2}$ & $\phi_b$ &      21.8 \\
 $n_c$ & $2.741267 \times 10^{+1}$ & $1.313261 \times 10^{+1}$ & $\phi_c$ &   125.4 \\
 $g_1$ & $2.534770 \times 10^{-2}$ & $1.420247 \times 10^{+4}$ & $\phi_1$ & 225.9 \\
 $g_2$ & $1.169675 \times 10^{-6}$ & $3.077778 \times 10^{+8}$ & $\phi_2$ & 145.8 \\ \hline
\end{tabular}
\end{center}
\end{table}

We performed a frequency analysis of the nominal orbital solution listed in Table~5 computed {over $10^5$\,yr.}
The orbits of the planets were integrated with the symplectic integrator SABA1064 of \citet{Farres_etal_2013}, using a step size of $5\times 10^{-3}$~yr and general relativity corrections.
The fundamental frequencies of the systems are the mean motions $n_b$ and $n_c$, and the two secular frequencies of the pericenters $g_1$ and $g_2$ (Table\,\ref{Tdyn1}).

To present the solution in a clearer way, it is useful to make a linear change of variables into eccentricity proper modes \citep[see][]{Laskar_1990}. 
In the present case, the linear transformation is numerically obtained with the frequency analysis of the solutions because of the high eccentricity
of the outer planet.
Using the classical complex notation $z_\kp = e_\kp \mathrm{e}^{i \varpi_\kp}$,
for $\kp = b, c$, the linear Laplace-Lagrange solution reads
\begin{equation}
\left(\begin{array}{c} z_b\\ z_c \end{array}\right)= 
\left(\begin{array}{rrr} 
0.058828 &  0.000961 \\
0.000000 &  0.811418 \\
\end{array}\right)
\left(\begin{array}{c} u_1\\ u_2 \end{array}\right) \ .
\label{eq.lape}
\end{equation}
To good approximation, the two proper modes $u_\lp$ (with $\lp = 1, 2$) are given by $u_\lp \approx  \mathrm{e}^{i (g_\lp t +\phi_\lp)}$, where $g_\lp$ and $\phi_\lp$ are listed in Table\,\ref{Tdyn1}.
Because the outer planet is much more massive than the inner one, there is almost no effect of the inner planet on the outer orbit: $z_c \approx 0.811\,u_2$, and thus $e_c = |z_c| \approx 0.811 = cte$ (Eq.\,(\ref{eq.lape})).

Equation~\ref{eq.lape} usually provides a good approximation for the long-term evolution of the eccentricities. 
In Fig.~\ref{Fdyn2} we plot the eccentricity evolution of the inner orbit with initial conditions from Table~5. 
Simultaneously, we plot its evolution given by the above secular, linear approximation ($e_b = |z_b|$). 
We see there is a good agreement with the numerical solution, that is, the eccentricity behaviour is described well by the secular approximation (Eq.\,\ref{eq.lape}).
The eccentricity variations result from the perturbation of the outer planet, but they are very limited ($0.05786 < e_b < 0.05980$).
These variations are driven mostly by the secular frequency $g_1$, with a period of approximately 14~kyr (Table\,\ref{Tdyn1}). 

\begin{figure}
    \includegraphics*[width=\columnwidth]{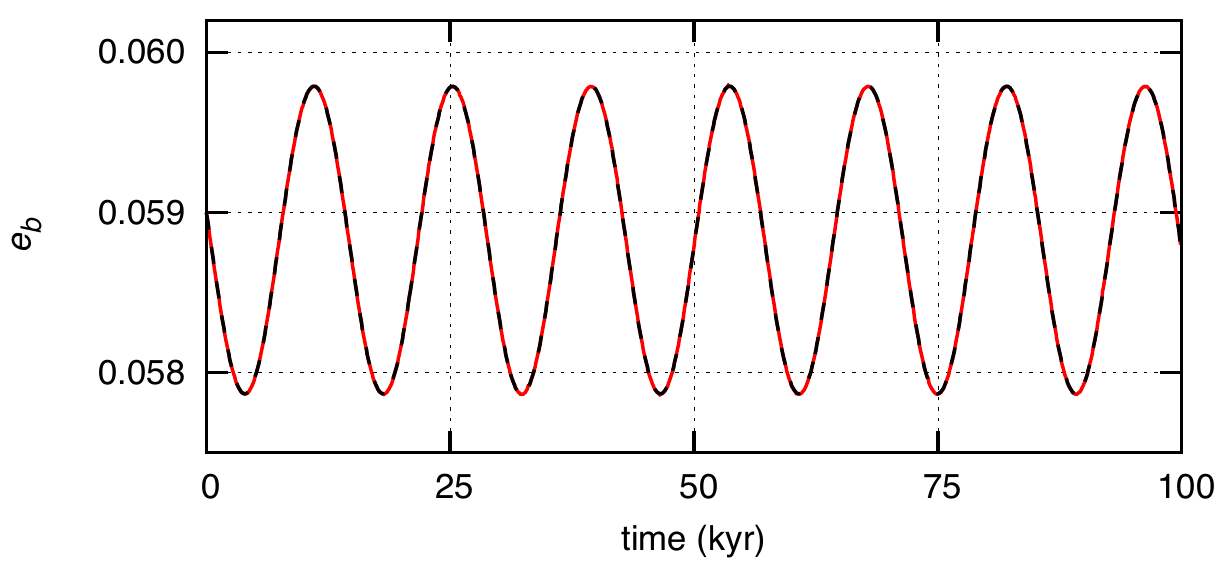} \\    
  \caption{Evolution of the eccentricity of the inner planet with time, starting with the orbital solution from Table~5. The red line gives the complete numerical solution, while the black dashed curve is obtained with the linear approximation (Eq.\,(\ref{eq.lape})). 
   \label{Fdyn2}}   
\end{figure}

\subsection{Tidal evolution}

The inner planet is very close to the star, and it is therefore expected to undergo strong tidal dissipation that slowly damps the eccentricity close to zero \citep[e.g.][]{Hut_1980, Correia_2009}.
The best-fit determination for the inner planet eccentricity is $e_b = 0.059 \pm 0.036$ (Table~5).
This value is still compatible with zero, but the question arises
whether we measure some residual equilibrium value resulting from the perturbations of the outer massive companion \citep{Mardling_2007, Laskar_etal_2012}.

The linear approximation from the previous section (Eq.\,\ref{eq.lape}) provides a direct estimate for the final equilibrium eccentricity.
In presence of tides, the proper modes are damped following an exponential decay law $|u_\lp| = \mathrm{e}^{-\gamma_\lp t}$, where $\gamma_\lp$ depends on the tidal dissipation \citep{Laskar_etal_2012}.
Assuming that dissipation occurs only on the inner body, we obtain $\gamma_1 \gg \gamma_2$ and $\gamma_2 \approx 0$. 
Therefore, after some time, the amplitude $|u_1|$ is fully damped, while $|u_2| \approx 1$ is almost unchanged.
From Eq.~\ref{eq.lape} we therefore have $e_b = |z_b| = |0.058828\, u_1 + 0.000961\, u_2| \approx 0.000961$.

To test this scenario, we integrated the system from Table~5 taking tidal dissipation in the inner body and general relativity corrections into account.
We adopted the same model as in Sect.~6.3 of \citet{Bonfils_etal_2013} with $R = 28,000$~km, $k_2 = 0.5$, and $\Delta t = 100$~s.
In Fig.~\ref{Ftides} we show the eccentricity evolution until it reaches its equilibrium value.
It stabilises around $0.000964$, which agrees with the prediction from the secular model (Eq.\,\ref{eq.lape}).
In our model we did not consider the rotational and tidal deformation of the planet, which would decrease the equilibrium eccentricity
even more \citep{Laskar_etal_2012}.
We therefore conclude that the currently observed value $e_b = 0.059$ is most likely overestimated.

\begin{figure}
    \includegraphics*[width=\columnwidth]{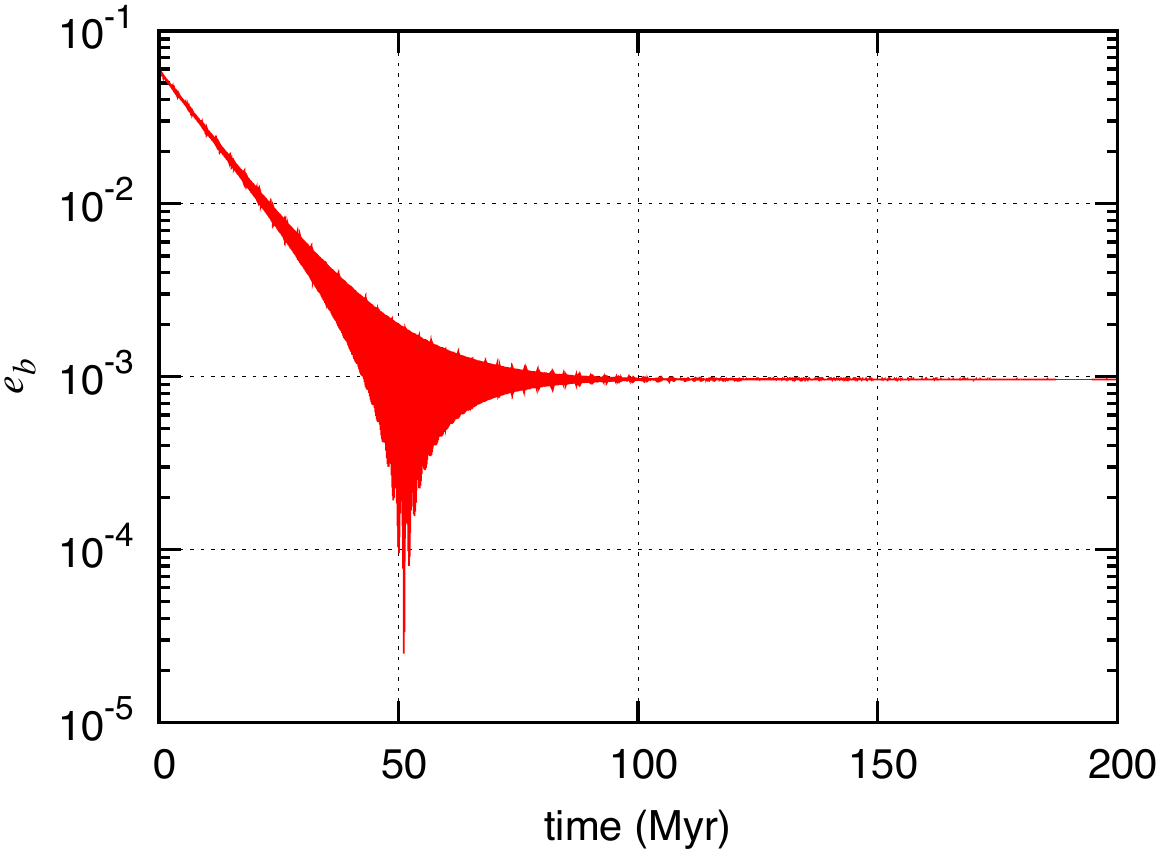} \\    
  \caption{Long-term evolution of the eccentricity of the inner planet with time, starting with the orbital solution from Table~5 and including tidal effects. The eccentricity stabilises around $e_b=0.00096,$ in perfect agreement with the prediction from the {secular linear model} (Eq.\,(\ref{eq.lape})). 
   \label{Ftides}}   
\end{figure}

}

\section{Discussion}
\label{sec:conclusions}

We presented the discovery of a massive giant planet in a long-period highly eccentric orbit
around HD\,219828, a G0IV metal-rich star for which a hot-neptune (HD\,219828\,b) was announced in \citet[][]{Melo-2007}.
With a minimum mass of $\sim$15\,M$_{Jup}$, HD\,219828\,c is
intermediate between the giant planet and brown-dwarf classes.
The data also allowed us to fully confirm the orbital parameters and mass for the shorter period
low-mass companion.

\subsection{Statistics and the formation of hot neptunes}

The system discovered around HD\,219828 is of interest for several reasons. The first and most evident reason is
the fact that {the two planets this system consists of are at the high- and low-mass ends of the mass distribution for
planet-mass companions to solar-type stars.} Studying it therefore may provide important clues to the formation of
short-period low-mass planets, as well as about the transition
from giant planets to brown dwarfs.

\begin{figure}[t]
\begin{center}
\begin{tabular}{c}
\includegraphics[angle=0,width=1\linewidth]{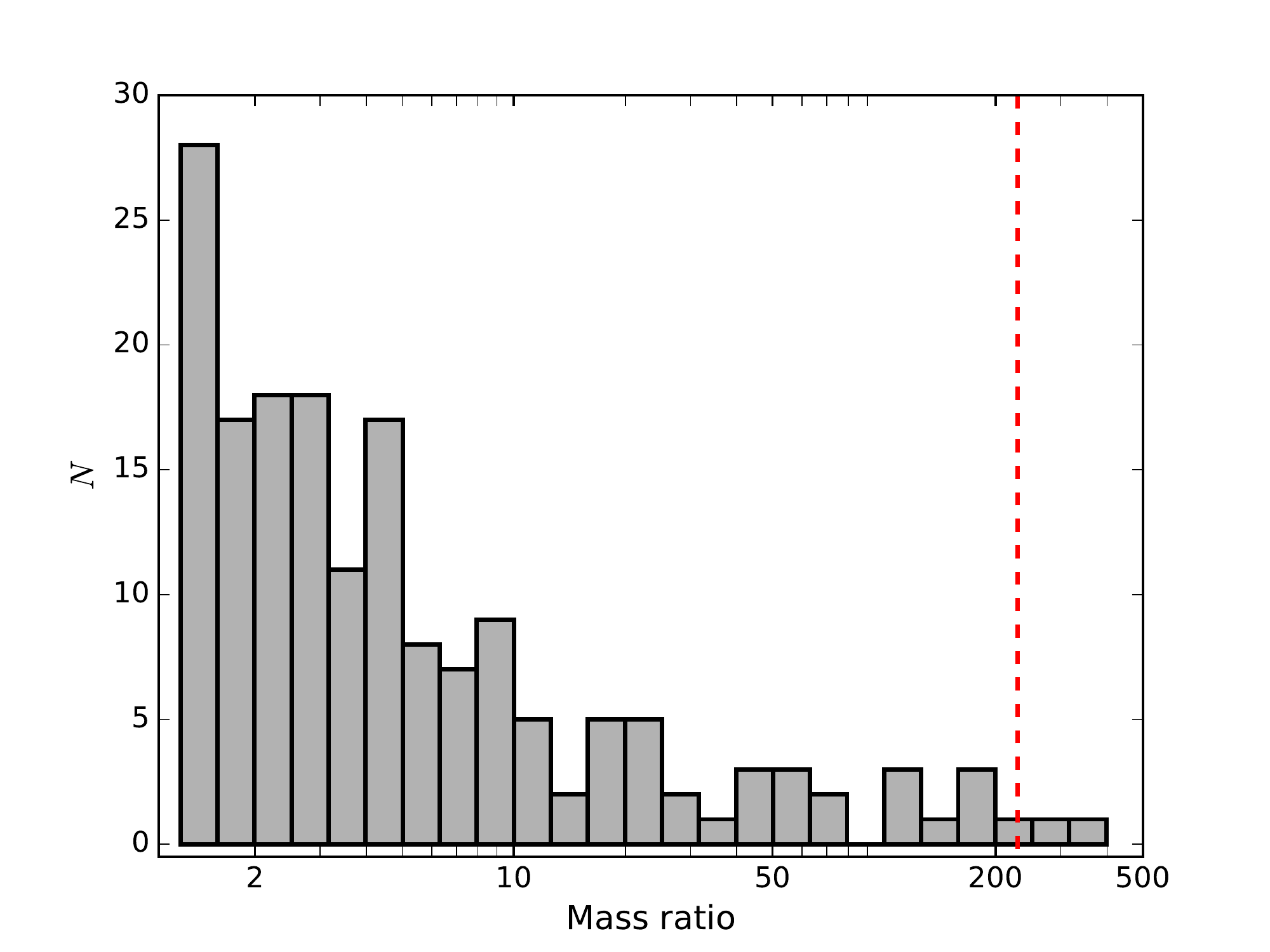}\\
\end{tabular}
\end{center}
\vspace{-0.5cm}
\caption{Distribution of mass ratios for all multi-planet systems listed in exoplanet.eu.}
\label{fig:massratio}
\end{figure}

By selecting all the planets discovered using the radial-velocity or transit 
methods in exoplanet.eu \citep[][]{Schneider-2011}, we conclude that the system orbiting HD\,219828
presents one of the highest mass ratios detected so far: assuming that the orbits are co-planar, m$_c$/m$_b$$\sim$229.
The peculiar mass ratio of this system is well illustrated in Fig.\,\ref{fig:massratio}, where we plot the mass ratio distribution
of all multi-planet systems listed in exoplanet.eu. For systems with three or more planets, only the higher mass
ratio is used for the histogram ($m_{higher}/m_{lower}$). The figure shows that most multi-planet systems have mass ratios below 10,
and only very few cases present mass ratios above 200. Except
for HD\,219828, other systems with a mass ratio above 200 include our solar system (not included in the plot), with a Jupiter-to-Mercury mass ratio of almost 4000\footnote{We note, however, that a planet like Mercury, as an exoplanet, is still beyond our detection capabilities.}, GJ\,676\,A, an M-dwarf where \citet[][]{Anglada-Escude-2012b} announced the presence of 2+2 super-Earths+jovian planets \citep[see also][]{Forveille-2011,Bonfils-2013} with the highest mass ratio of 353, \object{Kepler-94}, with a hot neptune and a moderately long-period (P$\sim$820\,days) jovian companion \citep[][a mass ratio of 288, assuming co-planar orbits]{Marcy-2014}, and \object{Kepler-454}, a system composed of a short-period transiting super-Earth, one jovian gas giant with a period of 524\,days, and an additional possible longer period gas giant \citep[][mass ratio of 207 assuming co-planar orbits]{Gettel-2016}. We note, however, that at least one of the low-mass planets in the Gl\,676\,A system has recently been disputed \citep[e.g.][]{Mascareno-2015}, 
and for planet c, the higher mass planet in the system, the only evidence presented in \citet[][]{Anglada-Escude-2012b} comes from a poorly constrained 
long-term trend in the data.

It is currently being discussed how often hot Jupiters are found in systems with additional longer
period companions. Recent results using direct imaging and radial-velocity surveys suggest that such companions are frequent
around hot-jupiter hosts: in about 50\% of the cases evidence exists for additional companions in the range 1-20 M$_{Jup}$ and with an orbital 
distance of 5-20 AU \citep[e.g.][]{Bryan-2016}. This rate is similar to the multiple star rate found in field stars \citep[][]{Raghavan-2010}.
The presence or absence of companions in longer period orbits may be key to understanding the processes of planet migration: disk induced \citep[e.g.][]{Lin-1996}, or dynamically induced (\citet[e.g.][]{Nagasawa-2008}. 
To our knowledge, however, a similar study has not been conducted for lower mass short-period planets.

We therefore searched in exoplanet.eu for all systems composed of a hot-neptune or super-earth and a longer
period massive companion (a giant planet) detected using the radial-velocity method. We did not include planets
detected by other methods to ensure that no strong bias was introduced (e.g. when the transit method
is strongly biased towards short-period planets). We found 40 stars with hot jupiters, 4 (or 10$^{+7}_{-3}$\%)\footnote{Error estimated following a binomial distribution.} 
of them are in multiple systems, all of them with high-mass planet companions (>0.1\,M$_{Jup}$). 
On the other hand, we found 38 stars with hot neptunes, 28 (or 74$^{+6}_{-8}$\%) of them in multiple systems: 
16 (42$^{+8}_{-7}$\%) have giant-planet companions (>0.1\,M$_{Jup}$) and 12 (32$^{+8}_{-6}$\%) have additional 
low-mass planets (neptunes or super-Earths). This strongly suggests that the frequency of companions, in particular longer period giant planets, 
in systems of hot neptunes is significantly higher than the one found for the hot-jupiter planets. This result may be a telltale sign that the
formation and evolution of hot neptunes follows a path different to that of their higher mass counterpars, the hot jupiters.

\subsection{A system full of potential}

The interest in the HD\,219828 system is also further enhanced because its high mass, the long orbital period, and the very high eccentric orbit of HD\,219828\,c (it is among the top 10 
known exoplanets with higher eccentricity), together with its distance to the Sun (78\,pc), make it a prime target for Gaia.
The relative semi-major axis of the orbital motion of HD\,219828\,c is $\sim$5.96\,AU, and the semi-major axis of the corresponding barycentric 
orbit of HD\,219828 is $\sim$0.07\,AU. This latter value corresponds to about 0.9\,mas at the distance of 78\,pc.
This signal is much stronger than the expected measurement uncertainties and should be detectable with GAIA, even 
though the nominal 5 yr mission lifetime does not cover one complete orbit. Using the combined astrometric and radial-velocity
measurements, we will then be able to derive a precise mass for the companion and also for its orbital parameters, including its orientation with respect to the sky.

Furthermore, given the radius of the star and the orbital period of HD\,219828\,b, the probability
that this planet transits is about 17\%. The hot neptune HD219828\,b is very similar to the transiting planet \object{Kepler-8\,b} \citep[][]{Jenkins-2010}, which has 
a mass of 24.4$\pm$3.8 M$_\oplus$ , a period of $\sim$3.21\,days, and is transiting a G0IV star. Kepler-8~b has a radius of 
about 4.0 R$_\oplus$. With a mass of 18.8$\pm$2.2 M$_\oplus$ and a period of 4.2\,days, the transiting planet 
\object{HAT-P-26~b} \citep[][]{Hartman-2011} is also similar to HD219828~b. However, HAT-P-26~b transits a later-type 
main-sequence star. It radius has been measured to be 6.3\,R$_\oplus$. Assuming HD219828~b has a mass of 21.0 M$_\oplus$ 
, we can expect its radius to be in the range 4--6 R$_\oplus$. Given the stellar radius of 1.69\,R$_\odot$ and an inclination of 90$\degr$, 
the transit is expected to have a depth in the range 600 -- 1200 ppm. For the most favourable case, this signal could be detected from the 
ground \citep[][]{Delrez-2015}. It would otherwise be an easy case for space-based photometry with a dedicated observatory such 
as CHEOPS \citep[][]{Fortier-2014} {or TESS \citep[][]{Ricker-2010}.
}Assuming a linear ephemeris for the inner planet, we find that the transit epochs can be computed using the following equation:
\begin{equation} 
T_{transit} = 2456001.237 \pm 5.6\,10^{-2} + n\times 3.834887 \pm 9.6\,10^{-5}
.\end{equation}
The covariance matrix between the initial transit epoch ($T_0$) and the period ($P$) is
\begin{equation}
cov(T_{0}, P) = \left\|\begin{array}{cc}3.012\times10^{-3} & 2.494301\times10^{-6} \\2.494301\times10^{-6} & 9.32304943\times10^{-9}\end{array}\right\|
.\end{equation}
At the time of the launch of CHEOPS, the uncertainty on the transit time is therefore at the level of two hours within a 68.3\% probability.

Furthermore, assuming a $v\,\sin{i}$ of 2.9\,$km\,s^{-1}$ and the above assumptions, the Rossiter-McLaughlin effect is expected 
to have a semi-amplitude that ranges between 1.4\,$m\,s^{-1}$ and 2.6\,$m\,s^{-1}$ and could be detected with an instrument like HARPS. 
If such a signal is detected, and assuming Gaia will provide an
orbital inclination (with respect to the sky) for HD\,219828\,c, we will be able to derive the
relative inclinations of the two companions. Such a unique measurement may allow us to 
set relevant constraints on the formation models of short-period low-mass planets \citep[see e.g.][]{Moriarty-2015}.

The long-period planet HD\,219828\,c has a minimum mass that positions it between a giant planet 
and a brown dwarf (this value may be known precisely when Gaia data become available) and orbits a bright
nearby star. It has the potential of becoming a benchmark object for the study of atmospheres. 
The 2MASS H magnitude of HD\,219828 is 6.6 \citep[][]{Skrutskie-2006}. Assuming
masses of 15, 50, and 80 times the mass of Jupiter for HD\,219828\,c, an age
of 5\,Gyr, and the distance of 78\,pc, the COND models of \citet[][]{Baraffe-2003} tell us that its apparent
H magnitude would be around 26, 20, and 15, respectively. While for the lower mass case
the magnitude difference may be too high for a detection using direct imaging techniques,
a flux ratio of about 10$^{-4}$ is expected if the companion is close to the brown dwarf/stellar
mass border. An even higher value may be expected at longer wavelengths.
Such signals should be easily detectable with recent instruments such as SPHERE
or using a detailed spectroscopic analysis \citep[][]{Brogi-2012,Snellen-2014,Martins-2015}.
We note that the projected distance on the sky between the star and the planet is about 0.14\,arcsec at apastron.

The system orbiting the bright HD\,219828, with its extreme mass ratio and the high-mass giant planet/brown dwarf companion 
may thus become a benchmark system for future studies of the formation of planetary systems
and of planet atmospheres.

%----------------------------------------------------------------------------------------
%       AcknowledgementsPapersAndBooks
%----------------------------------------------------------------------------------------
\begin{acknowledgements}
We would like to thank G. Chauvin for the interesting discussions. 
This work was supported by Funda\c{c}\~ao para a Ci\^encia e a Tecnologia (FCT, Portugal), project ref. PTDC/FIS-AST/1526/2014, through national funds and by FEDER through COMPETE2020 (ref. POCI-01-0145-FEDER-016886), as well as through grant UID/FIS/04434/2013 (POCI-01-0145-FEDER-007672).
P.F., N.C.S., and S.G.S. also acknowledge the support from FCT through Investigador FCT contracts of reference IF/01037/2013, IF/00169/2012, and IF/00028/2014, respectively, and POPH/FSE (EC) by FEDER funding through the program ``Programa Operacional de Factores de Competitividade - COMPETE''. P.F. and S.G.S. further acknowledge support from FCT in the form of exploratory projects of reference IF/01037/2013CP1191/CT0001 and IF/00028/2014/CP1215/CT0002. A.S. is supported by the European Union under a Marie Curie Intra-European Fellowship for Career Development with reference FP7-PEOPLE-2013-IEF, number 627202. E.D.M and V.Zh.A. acknowledge the support from the FCT in the form of the grants SFRH/BPD/76606/2011 and SFRH/BPD/70574/2010, respectively. A.S. thanks the Laboratoire d'Astrophysique de Marseille for the support in computing resources. J.P.F. acknowledges support from FCT through grant reference SFRH/BD/93848/2013. J.R. acknowledges CONICYT/Becas Chile 72140583. We warmly thank the OHP staff for their support on the 1.93 m telescope. We acknowledge support from the ``conseil scientifique'' of the Observatory of Paris and CIDMA strategic project UID/MAT/04106/2013. We gratefully acknowledge the Programme National de Plan\'etologie (telescope time attribution and financial support) of CNRS/INSU. This work results within the collaboration of the COST Action TD 1308. This work made use of SIMBAD.

\end{acknowledgements}

%----------------------------------------------------------------------------------------
%       Bibliography
%----------------------------------------------------------------------------------------
\bibliographystyle{aa}
\bibliography{santos_bibliography}

%----------------------------------------------------------------------------------
%       Appendices
%----------------------------------------------------------------------------------
%\appendix

%\section{Flux received by the ring}  \label{app:flux}

% Here, we detail the computation of the flux $F_r(\phi)$ received by the

\end{document}